\documentclass[10pt, letterpaper, conference]{IEEEtran}
\usepackage[english]{babel}
\usepackage{amsmath,amssymb}

\usepackage{mathtools}
\usepackage{graphicx}
\usepackage{caption}
\usepackage{subcaption}
\usepackage{url}
\usepackage{xcolor}
\usepackage{makecell, array}
\usepackage{diagbox}

\usepackage{subscript}
\usepackage{multirow}
\usepackage{tabularx}
\usepackage[binary-units]{siunitx}
\AtBeginDocument{
\DeclareSIUnit\bit{b}
}
\usepackage[inline]{enumitem}
\usepackage{footmisc}
\usepackage{balance}
\usepackage{listings}
\usepackage{url}
\usepackage{breakurl}
\usepackage[breaklinks]{hyperref}

\usepackage{todonotes}

\usepackage[acronym]{glossaries}
\newacronym{arm}{Arm}{}
\glsdisablehyper

\newcommand{\ie}{\emph{i.e.}, }
\newcommand{\eg}{\emph{e.g.}, }

\begin{document}

\title{Where to Encode: A Performance Analysis of x86 and \acrshort{arm}-based Amazon EC2 Instances}
\author{Roland Math\'{a}, Dragi Kimovski, Anatoliy Zabrovskiy, $\dots$,  Christian Timmerer, Radu Prodan}
\author{\IEEEauthorblockN{Roland Math\'a\IEEEauthorrefmark{1}, Dragi Kimovski\IEEEauthorrefmark{1}, Anatoliy Zabrovskiy\IEEEauthorrefmark{1},
Christian Timmerer\IEEEauthorrefmark{1}\IEEEauthorrefmark{2}, 
Radu Prodan\IEEEauthorrefmark{1}
}
\IEEEauthorblockA{\IEEEauthorrefmark{1}
Institute of Information Technology (ITEC), University of Klagenfurt, Austria}
\IEEEauthorblockA{\IEEEauthorrefmark{2}
Bitmovin, Klagenfurt, Austria}
}

\maketitle

\begin{abstract}
Video streaming became an undivided part of the Internet. To efficiently utilise the limited network bandwidth it is essential to encode the video content. However, encoding is a computationally intensive task, involving high-performance resources provided by private infrastructures or public clouds.
Public clouds, such as Amazon EC2, provide a large portfolio of services and instances optimized for specific purposes and budgets. The majority of Amazon's instances use \texttt{x86} processors, such as Intel Xeon or AMD EPYC. However, following the recent trends in computer architecture, Amazon introduced \texttt{\acrshort{arm}}-based instances that promise up to $40\%$ better cost performance ratio than comparable \texttt{x86} instances for specific workloads. We evaluate in this paper the video encoding performance of \texttt{x86} and \texttt{\acrshort{arm}} instances of four instance families using the latest FFmpeg version and two video codecs.
We examine the impact of the encoding parameters, such as different presets and bitrates, on the time and cost for encoding. Our experiments reveal that \texttt{\acrshort{arm}} instances show high time and cost saving potential of up to $33.63\%$ for specific bitrates and presets, especially for the \texttt{x264} codec. However, the \texttt{x86} instances are more general and achieve low encoding times, regardless of the codec.
\end{abstract}
\begin{IEEEkeywords}
Amazon EC2, Arm instances, AVC, Cloud computing, FFmpeg, Graviton2, HEVC, Performance analysis, Video encoding.
\end{IEEEkeywords}

\textcolor{red}{\small 2021 IEEE.  Personal use of this material is permitted.  Permission from IEEE must be obtained for all other uses, in any current or future media, including reprinting/republishing this material for advertising or promotional purposes, creating new collective works, for resale or redistribution to servers or lists, or reuse of any copyrighted component of this work in other works.}

\section{Introduction}
\label{sec:introduction}
Multimedia streaming content~\cite{cisco:report}, such as live and on-demand video and audio streams, is responsible for most Internet traffic today.
Unfortunately, the Internet network connectivity can significantly change over time depending on many factors, such as client location, network congestion, or end-user device~\cite{Midoglu:2019,AdViSE:2017}. The widely-used \textit{HTTP Adaptive Streaming (HAS)} technology~\cite{Sodagar2011} encodes video content, divided into small segments, in multiple bitrate-resolution pairs to adapt to varying bandwidth fluctuations. Using HAS, a video segment reaches clients at different bitrate-resolution pairs for optimized quality of experience, depending on its network characteristics and technical capabilities (e.g., viewing device, video player)~\cite{Bentaleb:2019}. Additionally, a number of HAS implementations are codec-independent~\cite{Sodagar2011} and allow providers to choose from a set of codecs for video encoding, including Advanced Video Coding (AVC)~\cite{AVCOverview}, High-Efficiency Video Coding (HEVC)~\cite{HEVCOverview}, VP9~\cite{VP9Overview}, AOMedia Video 1 (AV1)~\cite{Chen:2018}, and Versatile Video Coding (VVC)~\cite{VVC2019}.

Creating segments of a single video encoded with different parameters (e.g., bitrate, resolution) for adaptive streaming is a computationally-intensive process that requires expensive high-performance computers or cheaper cloud resources rented on-demand~\cite{Koziri:2018,Clouds2019}.
For example, Amazon EC2 provides different instance families optimized for specific purposes~\cite{Li:2019}, such as the balanced general purpose \texttt{m} instances, the compute-optimized \texttt{c} instances, the memory-optimized \texttt{r} instances, and the burstable \texttt{t} instances.
To further optimize video encoding workloads at a convenient price, leading video encoding companies such as Bitmovin (\url{https://bitmovin.com}) combine on-demand instances with EC2 Spot instances for video encoding~\cite{Amazon:2017}. Nevertheless, choosing cloud instances for thousands of encoding tasks is critical and can strongly influence the costs for the video service providers~\cite{Amazon:2017,kimovski2021cloud}.

Recently, Amazon launched their second generation Graviton \acrshort{arm}-based processors. \textit{Graviton2} is a 64-core monolithic server chip that uses \acrshort{arm}’s new Neoverse N1 cores, which is a derivation of the mobile Cortex-A76 cores. With the second generation of Graviton, Amazon EC2 promises higher performance at a lower cost compared to conventional \texttt{x86} instances. According to Amazon, Graviton2 instances, such as \texttt{m6g}, \texttt{c6g}, \texttt{r6g} and \texttt{t4g}, provide up to $40\%$ better price performance over comparable instance types with Intel Xeon processor for video encoding tasks~\cite{AmazonGraviton2}. 

In this paper, we analyze the video encoding performance of EC2 instances based on the Graviton2 \texttt{\acrshort{arm}} and the \texttt{x86} processors using four instance families optimized for different purposes. As Graviton2 processors are relatively new and its software support is continuously improving, we conduct all experiments using FFmpeg~\cite{FFmpeg:2020} employing the following video codec implementations:
\paragraph{\texttt{x264} codec} (H.264/AVC compression format) supported by the majority of end-user devices on the market~\cite{Practical-Evaluation:2018}. 
\paragraph{\texttt{x265} codec} (H.265/HEVC compression format) for further in-depth analysis with the newest FFmpeg version 4.3.

The main contributions of our work are:
\begin{itemize}[leftmargin=*,align=left]
\item We evaluate the relative video encoding time and cost differences between and \texttt{x86} and \texttt{\acrshort{arm}} instance families;
\item We identify the fastest and cheapest instance types among all instances of the same processor architecture.
\item We provide a reference table indicating our recommendation for the fastest and cheapest encoding options
for each instance family, preset, and bitrate.
\end{itemize}

The paper is organized into six sections. Section~\ref{sec:relatedwork} discusses the related work.  Section~\ref{sec:testingmethodology} presents the evaluation methodology. Section~\ref{sec:implementation} describes the experimental design and evaluation scenarios, followed by the two codecs' experimental results in Section~\ref{sec:evaluationx264} and Section~\ref{sec:evaluationx265}. Section~\ref{sec:summary-and-discussion} summarizes our recommendations and discusses the results. Finally, Section~\ref{sec:conclusion} concludes the paper and outlines the future work.

\section{Related work}
\label{sec:relatedwork}
We review the state-of-the-art related to performance analysis and characterization of video encoding on cloud instances. 

\paragraph{Cloud performance analysis} Li \textit{et al.}~\cite{Li:2019} analyze the performance of heterogeneous cloud instances for encoding video streams and provide a model for quantifying the suitability of cloud instance types for various encoding tasks. Xiangbo \textit{et al.}~\cite{li2017cost} present a performance analysis for improved scheduling of encoding tasks on cloud instances, which reduces the cost for the streaming service providers by $85\%$. Timmerer \textit{et al.}~\cite{timmerer2015live} present a performance analysis of the Bitmovin encoding platform, developed atop the \mbox{MPEG-DASH} open standard over multiple cloud instances. Based on this analysis, the encoding platform utilizes appropriate cloud instances to increase the average media throughput without stalling during operation.

\paragraph{Processor architecture} Jiang \textit{et al.}~\cite{Jiang:2020} present a performance characterization of cloud instances based on the first generation of Graviton compared to a multitude of Intel Xeon processors. The work demonstrates that although the Graviton processor has the slowest encoding speed and scalability ratio, it provides cost savings of around $15\%$ for the same encoding performance. Federman \textit{et al.}~\cite{ferdman2012clearing} analyze the micro-architectural behavior of \texttt{x86} processors for a set of video streaming workloads based on the \texttt{x264} codec. They identify that video encoding suffers from a high amount of stalled instructions and cache misses, which leads to lower scalability and hindered execution. Magaki \textit{et al.}~\cite{magaki2016asic} evaluate GPU and FPGA-based clouds' performance for video encoding with the \texttt{x265} codec and propose an application-specific integrated circuit tailored for this task.

\paragraph{Large scale video encoding} Jiang \textit{et al.}~\cite{Jiang:2019} evaluate the performance of public cloud infrastructure for large-scale video encoding applications scaling up to one thousand virtual machines. Regarding video encoding benchmarks, Lottatini \textit{et al.}~\cite{Lottarini:2018} present a public suite tailored for cloud video services, which encompasses a set of representative videos and metrics that reflect user-perceived characteristics of the video streams, such as quality and encoding speed. 

\paragraph{Gap analysis} The presented research works are segregated, utilize a single codec, and exclusively consider user-specific metrics, such a perceived video quality. In this work, we complement the related approaches by characterizing the performance of modern \texttt{x86} and \texttt{\acrshort{arm}} architectures for a set of commonly used video codec implementations. Furthermore, we consider various video encoding presets (\eg as known in \texttt{x264}, \texttt{x265}) and provide a detailed analysis on the performance and the cost savings of using optimized instances in the cloud. Lastly, we examine the suitability of the \texttt{\acrshort{arm}}-based processors for performing video encoding tasks.
\section{Evaluation methodology}
\label{sec:testingmethodology}
This section presents the performance evaluation methodology comprising two phases.
\begin{enumerate*}
    \item \emph{Encoding data generation} describes the selection of representative video sequences, identifies video codecs, and selects encoding parameters.
    \item \emph{Instance selection and metric definition} describes the selection of cloud instances based on the processor micro-architecture and defines the relevant performance evaluation metrics.
\end{enumerate*}

\subsection{Encoding data generation}
Encoding data generation involves video sequence, video codec, and encoding parameter selection.

\subsubsection{Video sequence selection}
\label{sec:tisi}
We select video segments with a duration below \SI{10}{\second} according to HAS requirements and industry best practices~\cite{Lederer:2012}. The segment length is a key parameter in HAS, as each video segment starts with a random access point to enable adaptive and dynamic switching to other representations (bitrate -- resolution pairs) at segment boundaries~\cite{Bouzakaria2014}. In addition, the selected video segment must contain movements that exploit different features of the video coding algorithms. Therefore, we use spatial and temporal information metrics to select a video segment and identify the computation requirements for its encoding ~\cite{itu910.2008}.

\paragraph{Spatial Information (SI)} measures the spatial complexity of video frames through the physical position of an object in the frame and its spatial relationship to other objects:
\begin{equation}
SI = \max_{\forall F_n}\{\sigma\left[\mathit{Sobel}\left(F_n\right)\right]\},
\label{eq:SI} 
\end{equation}
where $F_n$ is the luminance component of the video frame at time instance $n$, $\sigma$ is the standard deviation across all the pixels in the $\mathit{Sobel}$ filter, and the $\max$ operator calculates the maximum standard deviation of all frames in a video sequence. 
A high SI value indicates complex spatial relations between multiple objects and higher differences between subsequent frames, which increases the complexity of the encoding tasks and leads to longer encoding times.   

\paragraph{Temporal Information (TI)} shows the amount of motion in a video content, calculated as the maximum standard deviation $\sigma$ of a motion difference function $M_{n}(i,j)$. This function represents the difference in luminance for two sequential frames $F_n$ and $F_{n-1}$ across all the pixels $(i,j)$:
\begin{align}
TI = &\max\{\sigma\left[M_n(i,j)\right]\};\\
M_{n}(i,j) &= F_n(i,j) - F_{n-1}(i,j),
\label{eq:TI}
\end{align}  
where $F_n(i,j)$ is the luminance of the frame pixel $(i,j)$ at time instance $n$ in the video sequence.
A high TI value indicates higher motion differences between the video segment frames, which requires more computational resources for performing the encoding tasks. 

\subsubsection{Video codec selection}
We identify \texttt{x264} and \texttt{x265}~\cite{Practical-Evaluation:2018} as the most widely spread codecs for executing video encoding tasks, deployed by more than $90\%$ of the video streaming industry~\cite{itu-t.2019}. The \texttt{x265} video codec typically requires more computing resources than \texttt{x264} but achieves a higher video quality for the same encoding parameters~\cite{Huangyuan:2014}.

\subsubsection{Encoding parameters selection}
We select $19$ bitrates from \SI{100}{\kilo\bit\per\second} to \SI{20}{\mega\bit\per\second} (see Table~\ref{tab:bitrateladder}) and nine encoding presets that define the quality to encoding speed ratio:
ultrafast, superfast, veryfast, faster, fast, medium (default preset), slow, slower, and veryslow~\cite{ffmpeg-doc}. 
A slower preset uses more features for the same bitrate, which leads to a relatively slower encoding speed and better video quality~\cite{Silveira:2017}. Similarly, faster presets produce lower video quality. According to the official FFmpeg video encoding guide ~, we do not use the placebo preset that does not provide a significant quality improvement compared to the veryslow preset according to the official FFmpeg video encoding guide~\cite{ffmpeg-doc}.

\begin{table}[t!]
\centering
\normalsize
\caption{Bitrate ladder (bitrate -- resolution pairs).}
\label{tab:bitrateladder}
\resizebox{\columnwidth}{!}{%
\begin{tabular}{|c|c|c||c|c|c|}
\hline
\emph{\#} & \emph{Bitrate [\SI{}{\kilo\bit\per\second}]} & \emph{Resolution} & \emph{\#} & \emph{Bitrate [\SI{}{\kilo\bit\per\second}]} & \emph{Resolution}\\ 
\hline 
1 & $100$ & $256{\times}144$ & 11 & $4300$ & $1920{\times}1080$\\
2 & $200$ & $320{\times}180$ & 12 & $5800$ & $1920{\times}1080$\\
3 & $240$ & $384{\times}216$ & 13 & $6500$ & $2560{\times}1440$\\
4 & $375$ & $384{\times}216$ & 14 & $7000$ & $2560{\times}1440$\\
5 & $550$ & $512{\times}288$ & 15 & $7500$ & $2560{\times}1440$\\
6 & $750$ & $640{\times}360$ & 16 & $8000$ & $3840{\times}2160$\\
7 & $1000$ & $768{\times}432$ & 17 & $12000$ & $3840{\times}2160$\\
8 & $1500$ & $1024{\times}576$ & 18 & $17000$ & $3840{\times}2160$\\
9 & $2300$ & $1280{\times}720$ & 19 & $20000$ & $3840{\times}2160$\\
10 & $3000$ & $1280{\times}720$ & & & \\   
\hline
\end{tabular}
}
\end{table}

\subsection{Instance selection and metric definition}

\subsubsection{Instance type selection} We selected instance types based on three commonly used processors for video encoding.
\paragraph{Intel Xeon Platinum} is a multi-purpose processor based on the latest extension of the \texttt{x86} architecture with the Advanced Vector Extension (AVX-512) instruction set. It is a $28$-core server chip that can execute $56$ concurrent threads.
\paragraph{AMD EPYC} is a multi-purpose processor based on the \texttt{x86} Zen architecture. For the majority of instance types, AWS provides the first generation EPYC processor with up to $32$ cores and $64$ concurrent threads. However, for the \texttt{c} instances, AWS provides the second-generation EPYC processors with up to $64$ cores and $128$ threads per server chip.

The \texttt{x86} processors of Intel and AMD represent $87\%$ of the cloud instances~\cite{arm-market-share} and the majority of personal computers.

\paragraph{Graviton2} is the second generation of Graviton \texttt{\acrshort{arm}} processors  recently released by Amazon. It is a 64-core monolithic server chip that uses \acrshort{arm}’s new Neoverse N1 cores, derived from the mobile Cortex-A76 cores.
The \texttt{\acrshort{arm}} processors dominate the mobile segment with a $90\%$ market share~\cite{arm-market-share}. Nevertheless, there is a trend~\cite{ARM2020} from leading companies, such as Apple and Amazon, for using \texttt{\acrshort{arm}}-based processors to power personal computers and cloud instances.

\paragraph{Instance families} Based on these processor architectures, we select four instance families from the Amazon EC2 cloud (see Table~\ref{tbl:InstanceTypes})~\cite{Li:2019}:
\begin{enumerate*}
    \item \textit{balanced general purpose} \texttt{m} instances,
    \item \textit{compute-optimized} \texttt{c} instances,
     \textit{memory-optimized} \texttt{r} instances, and
     \item \textit{burstable} \texttt{t} instances.
\end{enumerate*}
We selected eight vCPUs for all instance types, corresponding to the largest available sizes of \texttt{t3}, \texttt{t3a}, and \texttt{t4g} instances. For a fair comparison, we equalize the number of vCPUs. However, the memory size of different instance types might still differ due to their instance type definitions. Nonetheless, the carefully selected video segment's encoding does not exceed the smallest memory size in our set of instances.

\subsubsection{Evaluation metrics}
We compare the new \texttt{\acrshort{arm}} instances with the Intel and AMD based \texttt{x86} instances using three metrics.
\paragraph{Relative encoding time} $\Delta{t_{enc}}(V_{b,p})$ of a video segment $V_{b,p}$ with a bitrate $b$ and a preset $p$ is the normalized time difference of the \texttt{\acrshort{arm}} encoding time $t_{Arm}(V_{b,p})$ to the reference \texttt{x86} encoding time $t_{x86}(V_{b,p})$:
\begin{equation}
\Delta{t_{enc}}(V_{b,p}) = \frac{t_{Arm}(V_{b,p})-t_{x86}(V_{b,p})}{t_{x86}(V_{b,p})} \cdot 100.
\end{equation}
A positive relative encoding time indicates that the \texttt{\acrshort{arm}} instance is slower than the reference \texttt{x86} instance, while a negative value indicates that \texttt{\acrshort{arm}} is faster.

\begin{table*}[t]
\centering
\caption{Experimental Amazon EC2 Cloud instance types.}
\begin{tabular}{|@{ }c@{ }|@{ }c@{ }|@{ }r@{ }|@{ }r@{ }|@{ }c@{ }|@{ }c@{ }|@{ }c@{ }|@{ }c@{ }|@{ }l@{ }|}
\hline
 \textit{Instance type} & \textit{Architecture}& \textit{vCPUs} & \textit{Memory [\SI{}{\gibi\byte}]} & \textit{Storage [\SI{}{\gibi\byte}]} & \textit{Network} & \textit{Physical processor} & \textit{Clock [\SI{}{\giga\hertz}]}& Price[\SI{}{\$\per\hour}]\\
\hline
\texttt{m5.2xlarge}& \multirow{4}{*}{64-bit x86} & \multirow{4}{*}{8} & 32 & \multirow{4}{*}{EBS} & \multirow{3}{*}{$\le 10$ \SI{}{\giga\bit\per\second}} & 1st or 2nd generation & $\le 3.1$&0.383\\
\texttt{c5.2xlarge}& & & 32 & & & Intel Xeon Platinum 8000 & $\le 3.4$&0.34\\
\texttt{r5.2xlarge}& & & 64 & & & series (Skylake-SP or & $\le 3.1$&0.504\\
\texttt{t3.2xlarge}& & & 32 & & $\le 5$ \SI{}{\giga\bit\per\second}& Cascade Lake) & $\le 3.1$&0.3341\\
\hline
\texttt{m5a.2xlarge}& \multirow{4}{*}{64-bit x86} & \multirow{4}{*}{8} & 32 & \multirow{4}{*}{EBS} & \multirow{3}{*}{$\le 10$ \SI{}{\giga\bit\per\second}} & AMD EPYC 7000 series& $\le 2.5$&0.344\\
\texttt{c5a.2xlarge}& & & 16 & & & AMD EPYC 7002 (2nd gen.)& $\le 3.3$&0.308\\
\texttt{r5a.2xlarge}& & & 64 & & & AMD EPYC 7000 series& $\le 2.5$&0.452\\
\texttt{t3a.2xlarge}& & & 32 & & $\le 5$ \SI{}{\giga\bit\per\second}& AMD EPYC & $\le 2.5$&0.3008\\
\hline
\texttt{m6g.2xlarge} & \multirow{4}{*}{64-bit \acrshort{arm}}& \multirow{4}{*}{8} & 32 & \multirow{4}{*}{EBS} & \multirow{3}{*}{$\le 10$ \SI{}{\giga\bit\per\second}} & \multirow{4}{*}{AWS Graviton2} & \multirow{4}{*}{$\le 2.5$}&0.308\\
\texttt{c6g.2xlarge} & & & 16 & & & &&0.272\\
\texttt{r6g.2xlarge} & & & 64 & & & &&0.4032\\
\texttt{t4g.2xlarge} & & & 32 & &$\le 5$ \SI{}{\giga\bit\per\second} & &&0.2688\\
\hline
\end{tabular}
\label{tbl:InstanceTypes}
\end{table*}

\paragraph{Encoding cost} $c_{enc}(V_{b,q})$ of a video segment $V_{b,q}$ is the product between the instance price $c_{i}$ (in \$) per second and the segment encoding time $t(V_{b,q})$ in seconds:
\begin{equation}
c_{enc}(V_{b,q}) = c_{i}\cdot t(V_{b,q}),
\end{equation}
Although cloud providers typically charge for their on-demand instances on an hourly basis, we scale the price down to seconds for a more fine-grained encoding cost understanding.

\paragraph{Relative encoding cost} $\Delta{c_{enc}}(V_{b,p})$ for a video segment $V_{b,p}$ with a bitrate $b$ and a preset $p$ is the normalized difference of the \texttt{\acrshort{arm}} encoding cost $c_{Arm}(V_{b,p})$ to the reference \texttt{x86} encoding cost $c_{x86}(V_{b,p})$:
\begin{equation}
\Delta{c_{enc}}(V_{b,p}) = \frac{c_{Arm}(V_{b,p})-c_{x86}(V_{b,p})}{c_{x86}(V_{b,p})} \cdot 100.
\end{equation}
Similarly to relative encoding time, a positive relative encoding cost indicates that \texttt{\acrshort{arm}} is more costly than the reference \texttt{x86} instance, while a negative value indicates that \texttt{\acrshort{arm}} is cheaper.

\section{Experimental design}
\label{sec:implementation}
This section describes the implementation and evaluation scenarios of the encoding methodology.

\subsection{Video sequence selection}
We selected for encoding a two-second segment from a computer-animated movie~\cite{Zabrovskiy:2018} with high TI and SI metrics~\cite{SITI:software}, as described in Section \ref{sec:testingmethodology}.
The selected video segment's TI metric has a value of $22$, which is three times higher than the average TI value ($8.2$) of the other movie segments. This implies high motion between the frames.
In turn, the SI metric of this segment has a value of $18.1$, which is slightly above the average SI value ($16.3$) of all movie segments, which implies higher spatial complexity.

\subsection{Encoding software}
We perform the encoding for \texttt{AVC} and \texttt{HEVC} video codecs using the \texttt{libx264} (short \texttt{x264}) and \texttt{libx265} (short \texttt{x265}) FFmpeg~\cite{FFmpeg:2020} library, as follows:
\begin{lstlisting}[language=bash,basicstyle=\ttfamily]
ffmpeg -y -i sintel_2sec0030.y4m 
  -r {fps} -vf scale={WxH} format=yuv420p
  -c:v {libx264, libx265}
  -preset {preset} -b {bitrate}
  output.mp4
\end{lstlisting}
\paragraph{fps} is the number of video frames per second, which is the same as in the original video segment (\ie $24$);
\paragraph{W${\times}$H} specifies the width \texttt{W} and height \texttt{H} (in pixels) of the encoded video segment (see Table~\ref{tab:bitrateladder}); 
\paragraph{preset} defines the encoding preset (\eg fast, slow);
\paragraph{bitrate} in a encoded video file  measured in \SI{}{\kilo\bit\per\second}, as defined in Table~\ref{tab:bitrateladder}.

We used the latest FFmpeg version 4.3 which uses the \texttt{libx265} version 3.4 with Huawei enhancements for faster \texttt{\acrshort{arm}} encoding~\cite{x265:Release-Notes}.
We carefully select the number of encoding threads equal to the number of available vCPUs of each instance type.
We run all instances in the US East Amazon EC2 region with two default Ubuntu Server 18.04 LTS images that include Python packages: \texttt{ami-0ac80df6eff0e70b5} for 64-bit \texttt{x86} instances, and \texttt{ami-0d221091ef7082bcf} for 64-bit \texttt{\acrshort{arm}} instances. We manually deployed the latest FFmpeg version 4.3 to replace the default 3.4.6 version of Ubuntu 18.04 LTS.

\subsection{Evaluation scenarios}
\label{seubsec:scenario}
We created two evaluation scenarios for each of the codec.
We repeated all experiments five times for a total of $10,260$ experiments for both \texttt{x264} and \texttt{x265} codec. This translates into a total cloud instance time of over \SI{260}{\hour}. We report the average values of the metrics described in the following paragraphs.

\subsubsection{Instance family} comparison uses two metrics:
\paragraph{Relative encoding time} difference $\Delta{t_{enc}}(V_{b,p})$ of each \texttt{x86} and \texttt{\acrshort{arm}} instance from the same instance family, as presented in Table~\ref{tbl:InstanceTypes}
(\ie \texttt{m}, \texttt{c}, \texttt{r}, \texttt{t});
\paragraph{Relative encoding cost} difference $\Delta{c_{enc}}(V_{b,p})$ across both processor architectures from the same instance family.

\subsubsection{Processor architecture} comparison uses two metrics:
\paragraph{Fastest encoding time} of \texttt{\acrshort{arm}} $t_{Arm}(V_{b,p})$ and \texttt{x86} instances $t_{x86}(V_{b,p})$ independent of the instance family.
\paragraph{Lowest encoding cost} $c_{enc}(V_{b,p})$ across all instance families from the same processor architecture.
\section{Evaluation results}
We present the encoding results that compare the \texttt{x86} and \texttt{\acrshort{arm}} instances for the \texttt{x264} and \texttt{x265} codecs, following the evaluation scenarios from Section~\ref{seubsec:scenario}. We use heatmaps to simplify the three-dimensional visualization of the encoding time and cost dependency on the encoding bitrates and presets.

\subsection{x264 Codec}
\label{sec:evaluationx264}

\subsubsection{Instance family}
\label{subsec:x264comparison-per-instance-family}

\begin{figure*}[!t]
\centering
\begin{subfigure}{\textwidth}
\includegraphics[width=\textwidth]{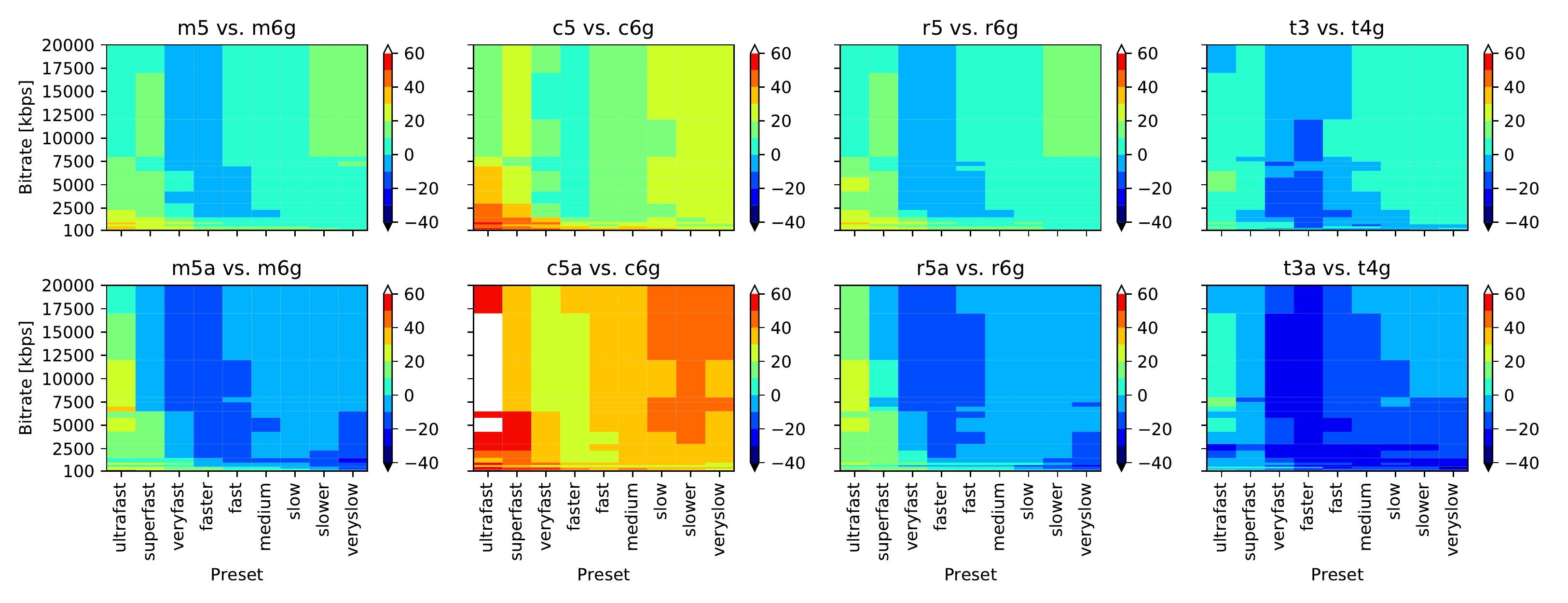}
\caption{Encoding times.}
\label{fig:x264-ffmpeg4-3-times}
\end{subfigure}
\begin{subfigure}{\textwidth}
\includegraphics[width=\textwidth]{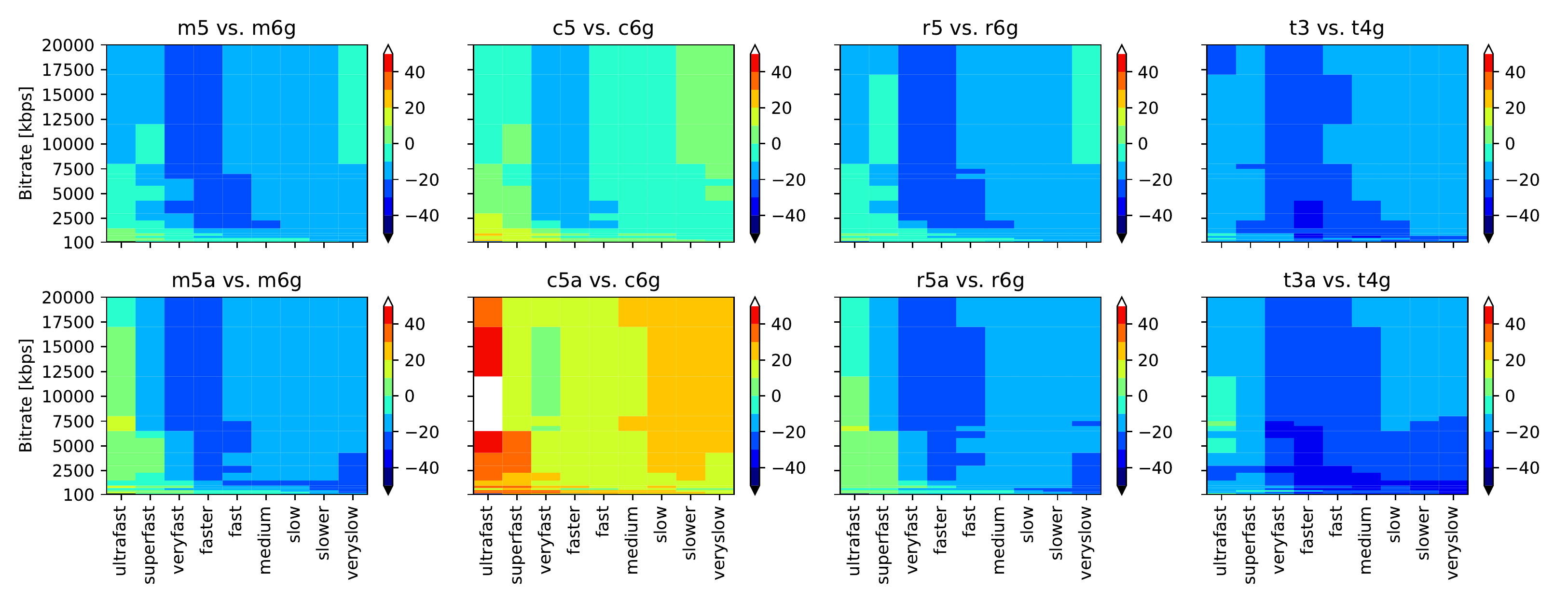}
\caption{Encoding costs.}
\label{fig:x264-ffmpeg4-3-costs}
\end{subfigure}
\caption{Relative encoding time and cost for four instance families with different presets, bitrates, and \texttt{x264} codec.}
\label{fig:x264-x86-arm-times-cost}
\end{figure*}

\paragraph{Relative encoding time}
Figure~\ref{fig:x264-ffmpeg4-3-times} displays the relative encoding time $\Delta{t_{enc}}(V_{b,p})$ between \texttt{\acrshort{arm}} and \texttt{x86} instances (Intel and AMD) with various encoding presets and bitrates.
We observe that the \texttt{x86 c} instances achieve for all presets and bitrates on average $30,78\%$ faster encoding times than the \texttt{\acrshort{arm} c6g} instances.
For the other instance families, the \texttt{x86} instances are faster than the \texttt{\acrshort{arm}} instances, primary for lower bitrates ($\leq$\SI{8}{\mega\bit\per\second}) and presets between ``ultrafast'' and ``veryfast''. 
The \texttt{\acrshort{arm}} instances \texttt{t4g}, \texttt{m6g} and \texttt{r6g} reveal faster encoding times than the Intel \texttt{x86} instances, 
especially for presets between ``very fast'' and ``fast'', and for bitrates higher than \SI{750}{\kilo\bit\per\second}. In particular, \texttt{m6g} and \texttt{r6g} achieve up to $7.51\%$, while \texttt{t4g} achieves up to $17.82\%$ faster encoding times than the Intel \texttt{x86} instances.
Considering the \texttt{x86} instances of AMD, the \texttt{\acrshort{arm}} instances \texttt{t4g}, \texttt{m6g} and \texttt{r6g} achieve faster encoding times especially for presets between ``very fast'' and ``very slow'', and for bitrates higher than \SI{750}{\kilo\bit\per\second}.

Overall, \texttt{t4g.2xlarge} shows the shortest relative encoding time due to the burstable behavior.

\paragraph{Relative encoding cost}
Beside relative encoding times, we closely analyze the relative encoding costs $\Delta{c_{enc}}(V_{b,p})$
with various encoding presets and bitrates (see Figure~\ref{fig:x264-ffmpeg4-3-costs}).
For \texttt{m}, \texttt{r}, and \texttt{t} instances, the \texttt{\acrshort{arm}} instances reveal the highest cost saving potential for presets between ``very fast'' and ``fast'', and bitrates over \SI{750}{\kilo\bit\per\second}.
For example, \texttt{t4g.2xlarge} shows on average  $19.67\%$, \texttt{r6g.2xlarge} $13.75\%$, and \texttt{m6g.2xlarge} $12.84\%$ lower encoding cost than the corresponding \texttt{x86} instances.
The \texttt{c} instances, especially the AMD based \texttt{c5a},  show highest cost saving potential for all presets and bitrates. The combination of lower cost and faster encoding performance than Intel based \texttt{c5} make the \texttt{c5a} to the first choice for the \texttt{x264} codec.

\paragraph{Recommendation} 
For fast encoding, we recommend using \texttt{x86} instances for bitrates lower than \SI{8}{\mega\bit\per\second} and presets between ``ultrafast'' and ``veryfast''. \texttt{\acrshort{arm}} instances are faster for bitrates higher than \SI{750}{\kilo\bit\per\second} and presets between ``very fast'' and ``fast''. For a low encoding cost, we recommend \texttt{\acrshort{arm}} over \texttt{x86} for \texttt{m}, \texttt{r}, and \texttt{t} instances. However, for \texttt{c} instances, we recommend \texttt{x86}, and especially the AMD based \texttt{c5a}, over \texttt{\acrshort{arm}}.

\subsubsection{Processor architecture}
\label{subsec:comp-per-proc-arch-x264}

\paragraph{\texttt{x86} encoding time}
The \texttt{c} instance family achieves the fastest encoding times in all experiments. Considering only Intel's \texttt{x86}, \texttt{c5} is on average  $10.44\%$ faster than other three Intel \texttt{x86} instances due to its higher clock speed.
However, the direct comparison of the Intel \texttt{c5}  and the AMD \texttt{c5a} reveal, that the AMD based \texttt{c} instance is on average $10.17\%$ faster. This performance difference is expected because the compute-optimized \texttt{c5a} instances use the second generation AMD EPYC processors, which are based on newer architecture and production processes than the Intel Xeon processors used in \texttt{c5}.

\begin{figure}[!t]
\centering
\begin{subfigure}{0.49\linewidth}
\includegraphics[width=\linewidth]{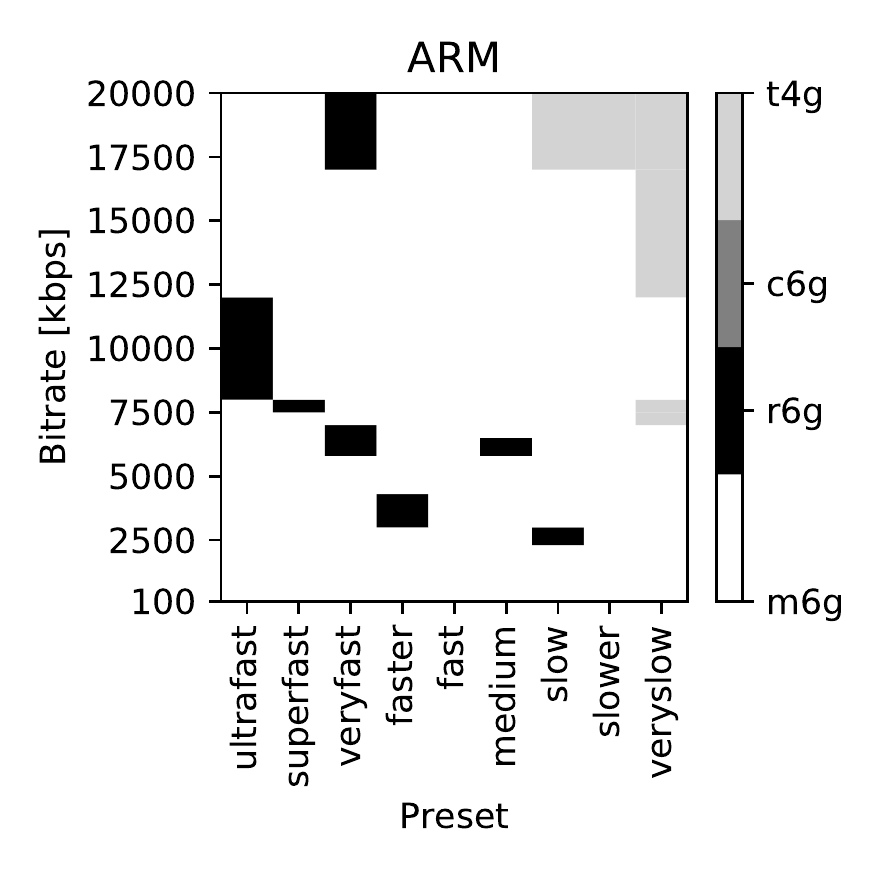}
\caption{Fastest.}
\label{fig:arm-fastest-x264}
\end{subfigure}
\begin{subfigure}{0.49\linewidth}
\includegraphics[width=\linewidth]{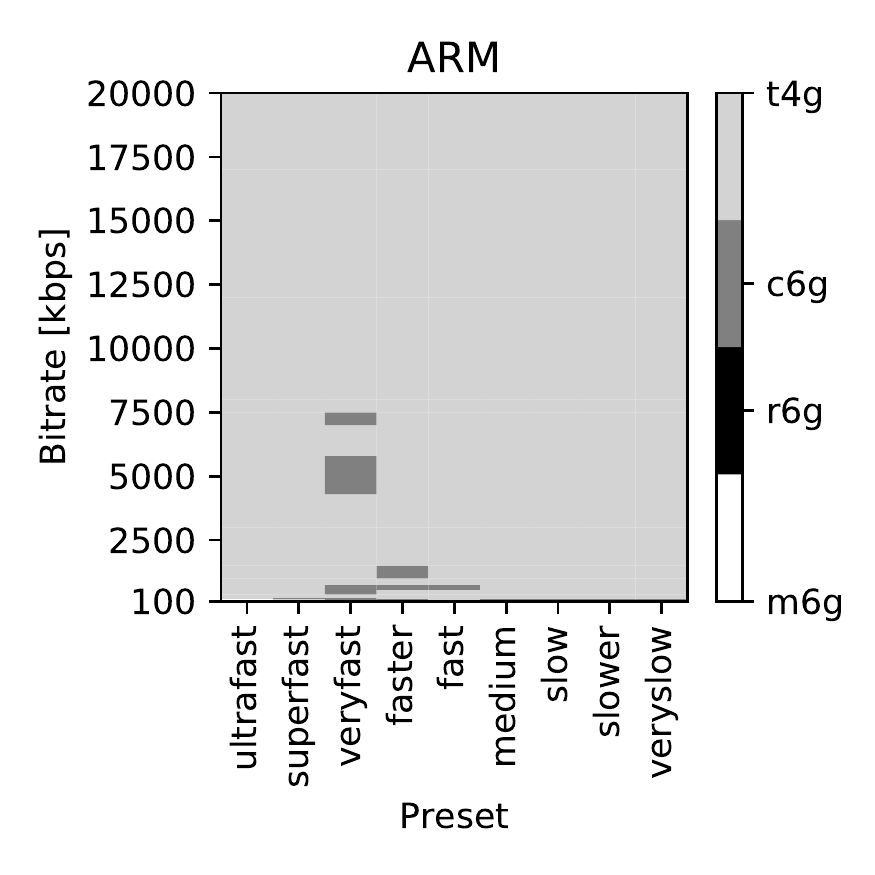}
\caption{Cheapest.}
\label{fig:arm-cheapest-x264}
\end{subfigure}
\caption{Fastest and cheapest encoding architecture per instance type for different presets, bitrates, and \texttt{x264} codec.}
\label{fig:arm-fastest-cheapest-x264}
\end{figure}

\paragraph{\texttt{\acrshort{arm}} encoding time} Figure~\ref{fig:arm-fastest-x264} shows the fastest encoding \texttt{\acrshort{arm}} instances. 
Overall, the encoding times among \texttt{\acrshort{arm}} instances are negligibly small with encoding time differences of less than $0.85\%$ on average.
Furthermore, Figure~\ref{fig:arm-fastest-x264} confirms that \texttt{m6g.2xlarge} is fastest in $90\%$ of the experiments with $1.15\%$ faster average encoding times. The \texttt{t4g.2xlarge} instance is fastest for bitrates higher than \SI{12}{\mega\bit\per\second} and presets between ``slower'' and ``veryslow''. The \texttt{r6g.2xlarge} instances are fastest in few cases with no detectable pattern, mainly for bitrates between \SI{2.5}{\mega\bit\per\second} and \SI{12}{\mega\bit\per\second}, and presets between ``ultrafast'' and ``slow''.

\paragraph{Encoding cost}
The \texttt{c} instance and especially the AMD-based \texttt{c5a} instance achieves the lowest \texttt{x86} encoding costs in all experiments thanks to its fast encoding times and low price compared to the other three \texttt{x86} instances.
Figure~\ref{fig:arm-cheapest-x264} shows the lowest encoding cost for \texttt{\acrshort{arm}} instances. In particular, the \texttt{t4g.2xlarge} instance achieves the lowest encoding costs, followed by \texttt{c6g.2xlarge}. The low encoding cost of \texttt{t4g.2xlarge} in the majority of experiments is due to its low price and burstable performance.

\paragraph{Recommendation}
Among \texttt{x86} instances, we recommend using \texttt{c} instances for fastest encoding and lowest cost. Among \texttt{c} instances, we recommend to prioritize \texttt{c5a} over \texttt{c5} instances, as \texttt{c5a} are based on latest processor architecture from AMD.
Among \texttt{\acrshort{arm}} instances, we recommend \texttt{m6g.2xlarge} for fastest encoding times, and \texttt{t4g.2xlarge} for lowest encoding cost.

\subsection{x265 codec}
\label{sec:evaluationx265}
To analyze the different codecs' impact, we repeat the same set of experiments using the \texttt{x265} codec.

\begin{figure*}[!t]
\centering
\begin{subfigure}{\textwidth}
\includegraphics[width=\textwidth]{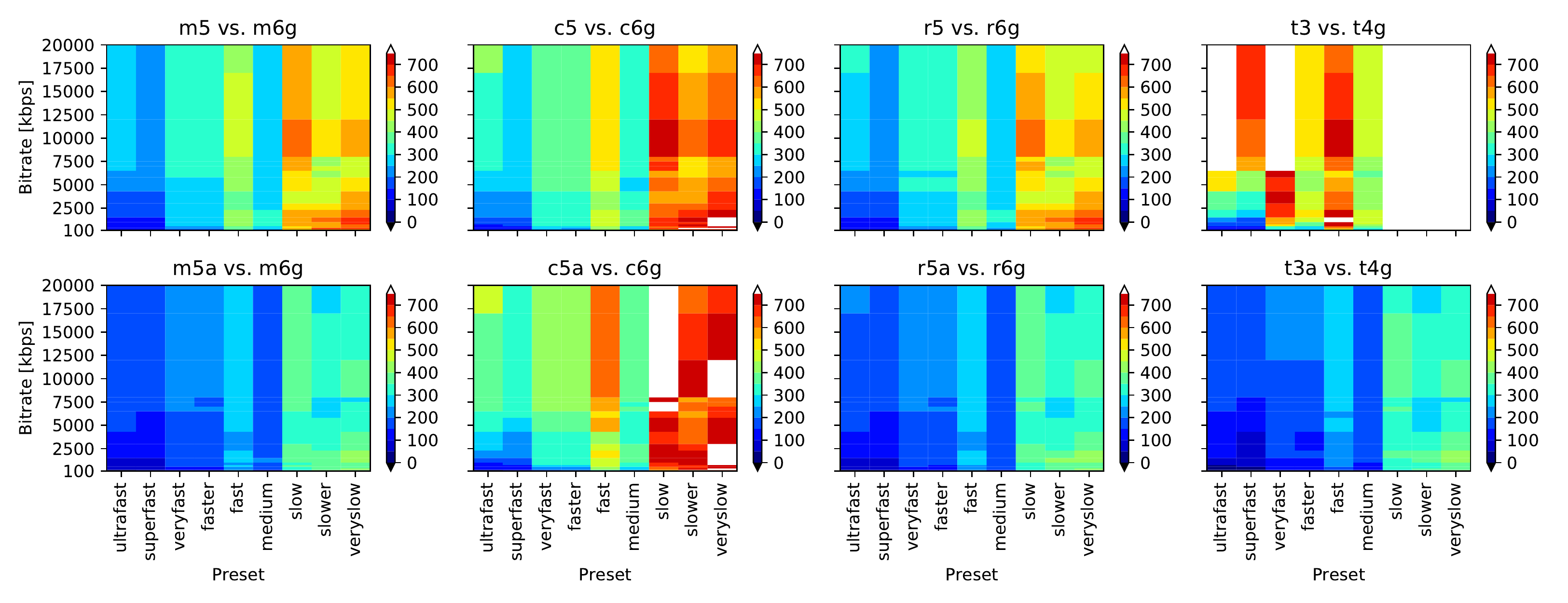}
\caption{Encoding times.}
\label{fig:x265-ffmpeg4-3-times}
\end{subfigure}
\begin{subfigure}{\textwidth}
\includegraphics[width=\textwidth]{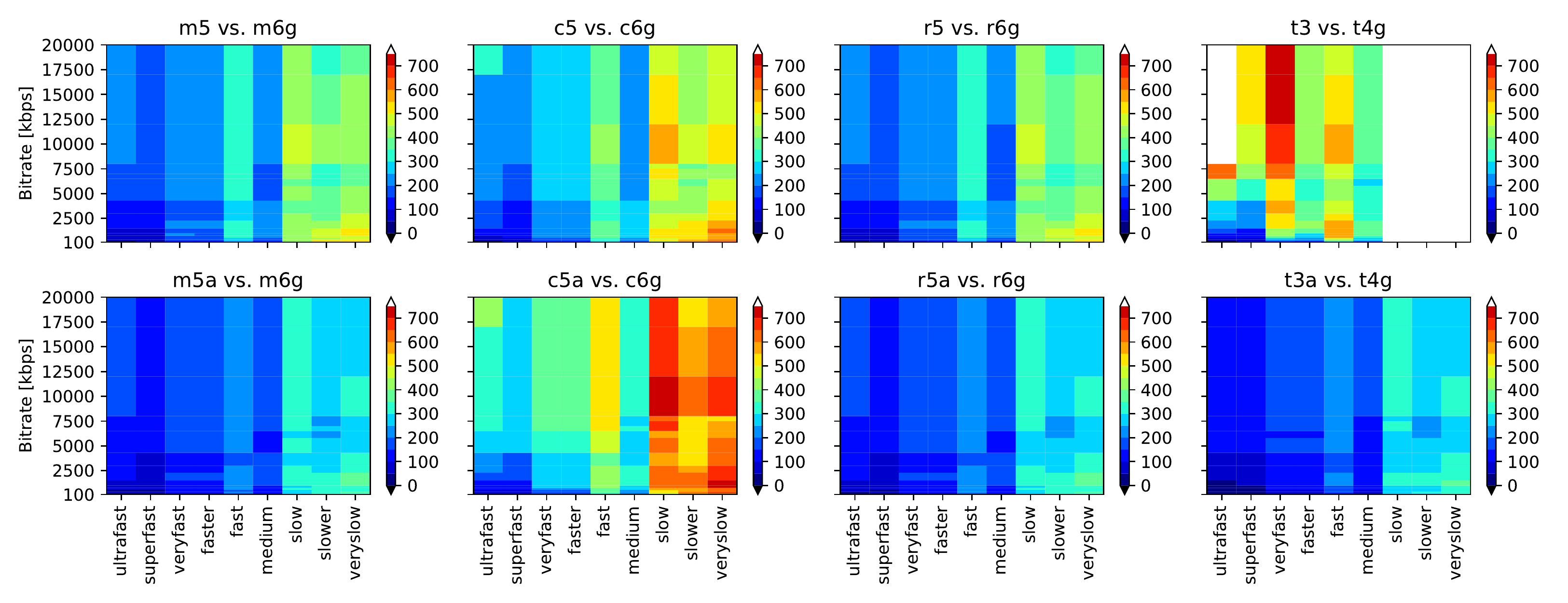}
\caption{Encoding costs.}
\label{fig:x265-ffmpeg4-3-costs}
\end{subfigure}
\caption{Relative encoding time and cost for four instance families with different presets, bitrates, and \texttt{x265} codec.}
\label{fig:x265-x86-arm-times-cost}
\end{figure*}

\subsubsection{Instance family}
\paragraph{Relative encoding time}
Figure~\ref{fig:x265-x86-arm-times-cost} reveals that the \texttt{\acrshort{arm}} instances are on average $329.42\%$ slower than the \texttt{x86} instances in all experiments. 
We observe higher relative encoding times of at least $279.92\%$ for presets between ``slow’’ and ``veryslow’’ and lower for presets between ``ultrafast’’ and ``superfast’’. In particular, \texttt{t4g.2xlarge} shows the smallest difference of $16.48\%$ for a bitrate of \SI{100}{\kilo\bit\per\second}, and \texttt{c5a} the highest differences of up to $847.67\%$ compared to \texttt{\acrshort{arm} c6g} instance for presets between ``slow’’ and ``veryslow’’.
The \texttt{m6g.2xlarge} and \texttt{r6g.2xlarge} instances achieve comparable results depending on the \texttt{x86} processor. We identify a higher average relative encoding time difference of $360.79\%$ for Intel based instances, and $229.76\%$ for AMD based instances.

In contrast to the \texttt{x264} codec, \texttt{\acrshort{arm}} instances show in all experiments slower relative encoding times. 
Moreover, we identify different areas in the heatmaps with small relative encoding time for both codecs. For example, the \texttt{x265} codec shows the smallest difference for areas with low bitrate and ``ultrafast’’ preset, where the \texttt{x264} codec shows highest $\Delta{t_{enc}}(V_{b,q})$.

\paragraph{Relative encoding cost}
For completeness, Figure~\ref{fig:x265-ffmpeg4-3-costs} depicts the relative encoding cost $\Delta{c_{enc}}(V_{b,p})$ between \texttt{x86} and \texttt{\acrshort{arm}} instances with various encoding presets and bitrates.
The \texttt{\acrshort{arm}} instances introduce $382.58\%$ higher encoding costs on average than the corresponding \texttt{x86} instances in all experiments. 

As \texttt{\acrshort{arm}} instances imply $31.66\%$ to $635.01\%$ higher encoding costs compared to Intel-based and $4.09\%$ to $736.90\%$ for AMD based\texttt{x86} instances. Therefore, we do not identify any cost-saving potential. We explain the higher costs due to the high relative encoding time, as presented in the previous section.

\paragraph{Recommendation} The experimental results confirm that the \texttt{x265} support in FFmpeg 4.3 is not yet optimized for \texttt{\acrshort{arm}} instances. We recommend using \texttt{x86} instances.

\subsubsection{Processor architecture}
\begin{figure}[!t]
\centering
\begin{subfigure}{0.49\linewidth}
\includegraphics[width=\linewidth]{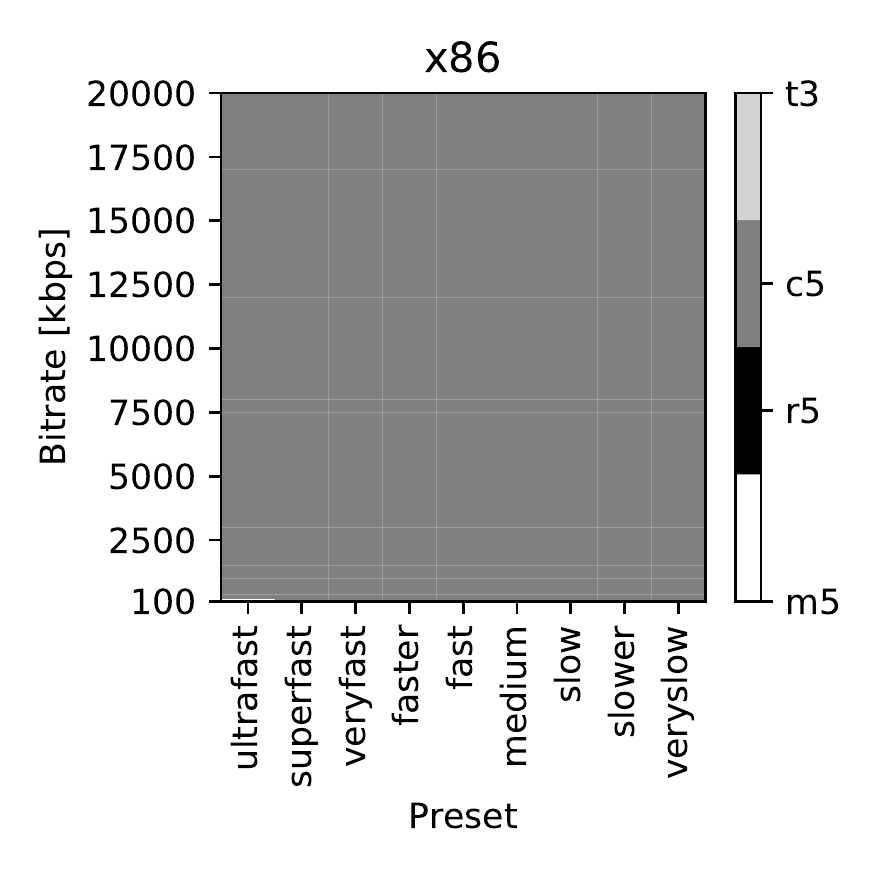}
\caption{\texttt{x86} instances.}
\label{fig:x86-x265-ffmpeg4-3}
\end{subfigure}
\begin{subfigure}{0.49\linewidth}
\includegraphics[width=\linewidth]{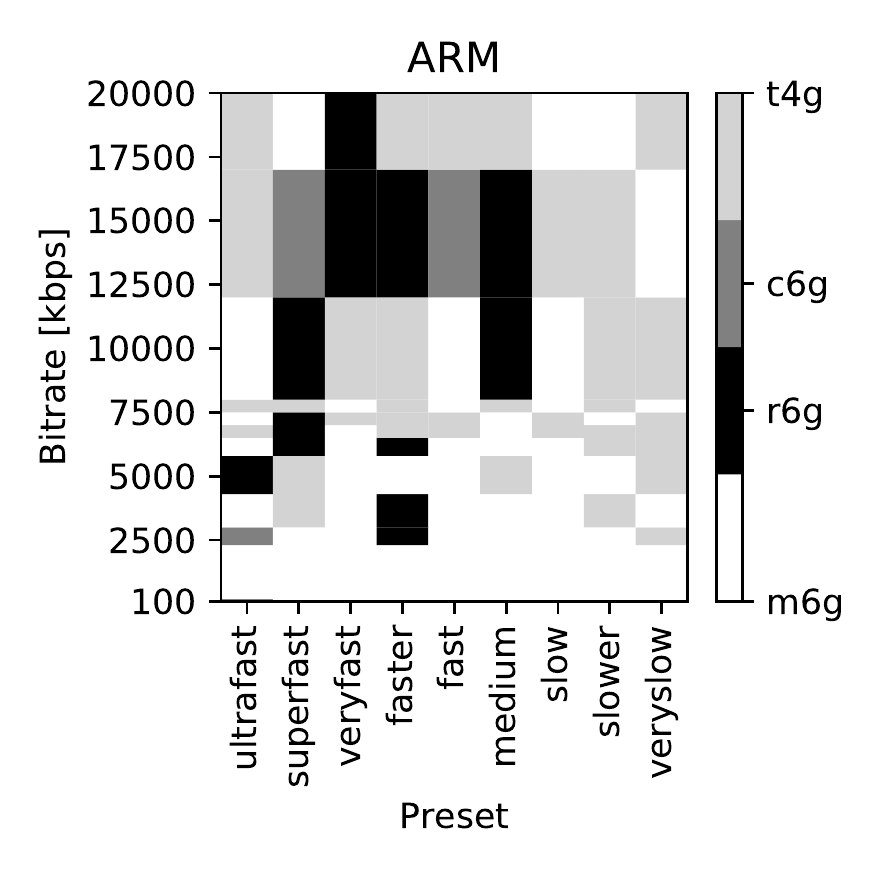}
\caption{\texttt{\acrshort{arm}} instances.}
\label{fig:arm-x265-ffmpeg4-3}
\end{subfigure}
\caption{Fastest encoding instance for different encoding presets, bitrates, and \texttt{x265} codec.}
\label{fig:fastest-x86-arm-x265}
\end{figure}

\paragraph{\texttt{x86} encoding time}
Figure~\ref{fig:x86-x265-ffmpeg4-3} shows that the \texttt{c} family delivers the fastest encoding time in all experiments on \texttt{x86} instances. Especially, the \texttt{c5a} instance with AMD processors is on average $15.65\%$ faster than all other \texttt{x86}. In particular,  \texttt{c5a} is on average $31.64\%$ faster than \texttt{t} instances, $27.19\%$ faster than \texttt{x86 m} and \texttt{r} instances, and $4.67\%$ faster than \texttt{c5} instances with Intel processor.


\paragraph{\texttt{\acrshort{arm}} encoding time}
Figure~\ref{fig:arm-x265-ffmpeg4-3} shows that \texttt{m6g.2xlarge} is fastest in $66.08\%$ of the experiments, followed by \texttt{t4g.2xlarge} in $23.39\%$.
Overall, the relative encoding time among the \texttt{\acrshort{arm}} instances is negligibly small and in the range of measurement noise. More precisely, \texttt{m6g.2xlarge} is on average only $0.21\%$, \texttt{t4g.2xlarge} $0.13\%$, \texttt{r6g.2xlarge} $0.23\%$, and \texttt{c6g.2xlarge} $0.07\%$ faster. 

\paragraph{Encoding cost}
For all experiments, \texttt{c5a.2xlarge} achieves lowest encoding costs among \texttt{x86} instances, and \texttt{t4g.2xlarge} among \texttt{\acrshort{arm}} instances.
Specifically, the \texttt{c5a} instance with AMD processors reveals on average $15.65\%$ lower cost than all other \texttt{x86} instances and $13.65\%$ lower average cost than the corresponding \texttt{c5} instance with Intel processors.
For comparison, the \texttt{t4g.2xlarge} reports only $1.34\%$ lower cost among all \texttt{\acrshort{arm}} instances.

\paragraph{Recommendation}
The \texttt{c5a.2xlarge} instance shows the fastest encoding times and the lowest encoding cost for \texttt{x86} instances in all experiments.
Among the \texttt{\acrshort{arm}} instances, \texttt{t4g.2xlarge} competes with \texttt{m6g.2xlarge} for the fastest encoding time. However, \texttt{t4g.2xlarge} achieves the lowest encoding cost in all experiments.

\section{Summary and Discussion}
\label{sec:summary-and-discussion}
\subsection{Summary}
\begin{figure*}[!t]
\centering
\begin{subfigure}{\textwidth}
\includegraphics[width=\textwidth]{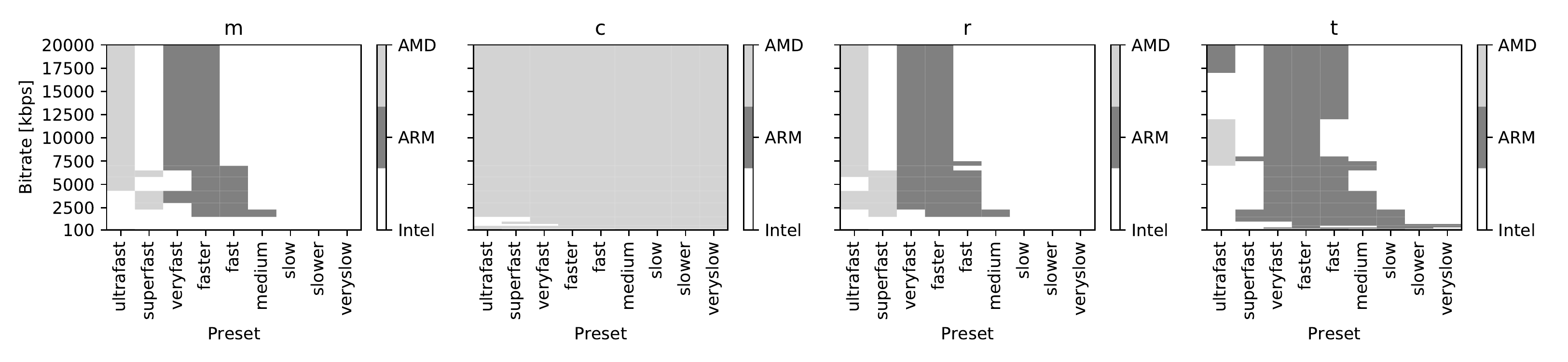}
\caption{Fastest encoding.}
\label{fig:x264-ffmpeg4-3-fastest}
\end{subfigure}
\begin{subfigure}{\textwidth}
\includegraphics[width=\textwidth]{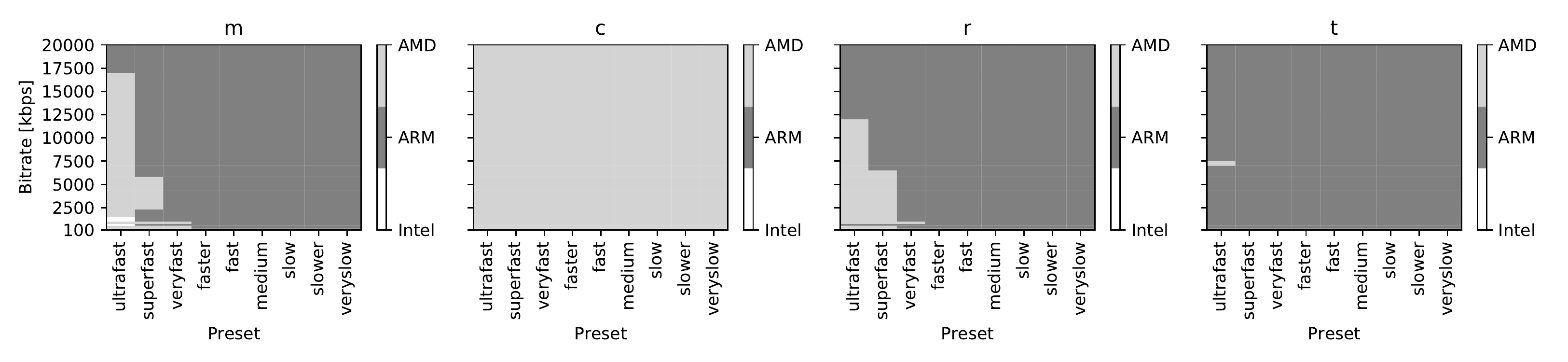}
\caption{Lowest encoding cost.}
\label{fig:x264-ffmpeg4-3-cheapest}
\end{subfigure}
\caption{Reference table for fastest encoding times and lowest cost for four instance families with different presets, bitrates, and \texttt{x264} codec.}
\label{fig:x264-summary-times-cost}
\end{figure*}

We summarize in this section our recommendations related to the video encoding time and cost for both codecs and processor architectures. In Figure \ref{fig:x264-summary-times-cost} we provide accurate visual reference tables indicating the fastest and cheapest options for all evaluated instance families, presets, and bitrates for the \texttt{x264} codec. As the results for \texttt{x265} codec report a clear best performing instance family for the fastest time and lowest cost, independently from the preset and bitrate, we describe these results only textually.

\subsubsection{Fast encoding}
\paragraph{\texttt{x264} codec} Overall, we recommend \texttt{x86} instances especially for bitrates lower than \SI{8}{\mega\bit\per\second} and presets between ``ultrafast'' and ``veryfast''. As depicted in Figure~\ref{fig:x264-ffmpeg4-3-fastest}, we recommend \texttt{\acrshort{arm}} instances mainly for bitrates higher than \SI{750}{\kilo\bit\per\second} and presets between ``very fast'' and ``fast''.
Among all \texttt{x86} instances, we recommend to use \texttt{c} instances and especially to prioritize the AMD based \texttt{c5a} over the Intel based \texttt{c5} instance.
Among all \texttt{\acrshort{arm}} instances,the \texttt{m6g.2xlarge} instance is the fastest, but with an advantage of only $1.15\%$ on average.

\paragraph{\texttt{x265} codec} We recommend \texttt{x86} instead of \texttt{\acrshort{arm}} instances, as the \texttt{x265} implementation in the latest FFmpeg version is not yet optimized for \texttt{\acrshort{arm}}. 
Among \texttt{x86} instances, we recommend the AMD based \texttt{c5a.2xlarge}, which is on average $4.67\%$ faster than the Intel based \texttt{c5.2xlarge} instance.
Among the \texttt{\acrshort{arm}} instances, we recommend \texttt{m6g.2xlarge} as it achieves the fastest encoding times in most experiments.  

\subsubsection{Low cost encoding}
\paragraph{\texttt{x264} codec}  Overall and as presented in Figure~\ref{fig:x264-ffmpeg4-3-cheapest}, we recommend \texttt{x86} instances for bitrates lower than \SI{8}{\mega\bit\per\second} and presets between ``ultrafast'' and ``superfast''.
Among all \texttt{x86} instances, we recommend to use \texttt{c} instances and to prioritize the AMD based \texttt{c5a} instance.
We recommend \texttt{\acrshort{arm}} instances for \texttt{m}, \texttt{r}, and \texttt{t} instances especially for presets between ``very fast'' and ``veryfast''. In particular, \texttt{\acrshort{arm}} instances reduce the encoding cost of up to $33.63\%$ compared to \texttt{x86}. 
We observe that \texttt{c5a} achieves the lowest encoding cost among \texttt{x86}, and \texttt{t4g.2xlarge} among \texttt{\acrshort{arm}} instances.

\paragraph{\texttt{x265} codec} We recommend using \texttt{x86} instances that generate on average $382.58\%$ lower encoding costs than \texttt{\acrshort{arm}} instances. The \texttt{c5a.2xlarge} instance achieves $15.65\%$ lower average cost among \texttt{x86} instances and on average $13.65\%$ lower cost than \texttt{c5.2xlarge} instances. Among \texttt{\acrshort{arm}} instances, the \texttt{t4g.2xlarge} reveals lowest cost in all experiments, but the advantage is just $1.34\%$.

\subsection{Discussion}
The pricing model of AWS ranks the instances in descending order with Intel \texttt{x86} instances with highest cost, followed by AMD based \texttt{x86} as middle cost, and \texttt{\acrshort{arm}} instances as low-cost alternatives. 
We observe that the instance performance primarily follows this pricing model. However, as cost and performance are two conflicting objectives, we also analysed the cost to performance ratio.
We revealed that for specific settings and codecs the low cost \texttt{\acrshort{arm}} instances can outperform faster and higher price \texttt{x86} instances. We explain this observation by the different instruction set architecture (ISA) of both processors types.

In particular, the \texttt{x86} CPUs use Complex Instruction Set Computing (CISC) while \texttt{\acrshort{arm}} uses Reduced Instruction Set Computing (RISC). With other words, the former uses more complex instructions with several cycles while \texttt{\acrshort{arm}} uses only one cycle to execute a single instruction. Consequently, we identified the software support for \texttt{\acrshort{arm}} instances as an important aspect that affects the performance. For example, we observed a high performance discrepancy between \texttt{\acrshort{arm}} and \texttt{x86} instances for the compute intensive \texttt{x265} codec. We explain the discrepancy through the fact that the FFmpeg 4.3 is not yet optimised for \texttt{\acrshort{arm}} instances.

Besides, the performance depends on the hardware allocation of the cloud provider. For example, the default allocation of Amazon EC2 uses hyper-threading that assigns on \texttt{x86} instances one logical hyper-threaded core to a vCPU (except \texttt{t2}). In contrast, the \texttt{\acrshort{arm}} instances assign one physical core to a vCPU~\cite{opt-cpu-options} in the absence of hyper-threading. This fact enables companies\footnote{\url{https://aws.amazon.com/ec2/graviton}}, such as Snap Inc reducing CPU utilization by roughly $10\%$, Honeycomb.io running $30\%$ fewer instances, and NextRoll saving up to $50\%$ cost compared to previous generation EC2 instances for their workloads.

Overall, with continuously improving software support and optimisation for \texttt{\acrshort{arm}}, we forecast a decreasing performance difference between \texttt{\acrshort{arm}} and \texttt{x86} instances which also applies for video encoding tasks.
\section{Conclusion}
\label{sec:conclusion}
In this paper, we provide a performance analysis for video encoding tasks of \texttt{\acrshort{arm}} and \texttt{x86} instances of four Amazon EC2 instance families with three different processors. We conducted a total of \num{20,520} experiments in two evaluation scenarios for the two most widely used video codecs.

Table~\ref{tbl:recomend} summarises our recommendations based on the evaluation results related to the encoding time and cost.

\begin{table}[t]
\scriptsize
\caption{Encoding recommendation summary.}
\centering
{
\resizebox{\columnwidth}{!}{
\begin{tabular}{|c|c|c|c|}
\hline
 \backslashbox{\textit{Codec}}{\textit{Goal}} &
 \textit{Fast encoding} & \textit{Low cost} \\
\hline
\textit{x264} & \makecell{\texttt{x86 (c5a)}\\\texttt{\acrshort{arm} (m6g)}} & \makecell{\texttt{\acrshort{arm} (t4g)}} \\
\hline
\textit{x265} & \makecell{\texttt{x86 (c5a)}} & \makecell{\texttt{x86 (c5a)}} \\
\hline
\end{tabular}
}}
\label{tbl:recomend}
\end{table}

The evaluation results clearly reveal that the \texttt{\acrshort{arm}} instances can achieve faster encoding times at a lower cost than the corresponding \texttt{x86} instances. However, the encoding performance of \texttt{\acrshort{arm}} and \texttt{x86} instances depends on many factors such as bitrate, preset, and codec. Independently from the codec, we show that the \texttt{x86} \texttt{c} instance and specifically the AMD based \texttt{c5a} instance achieves fast encoding times at a lower cost in most experiments, which makes it suitable for general encoding use.

In summary, regarding the measured encoding performance potential for specific encoding settings and the continuously improving support of \texttt{\acrshort{arm}} instances, we forecast a decreasing performance difference between \texttt{\acrshort{arm}} and \texttt{x86} instances.

In the future, we plan to extend our analysis with different emerging codec implementations and longer video segments to generalize our recommendations. Furthermore, we will perform an in-depth analysis of overprovisioning by evaluating \texttt{x86} instances with various CPU options~\cite{opt-cpu-options} for the fastest instance setup.



\balance
\bibliographystyle{plain}
\bibliography{bibliography}

\end{document}